\documentclass[a4paper,12pt]{jpconf}
\usepackage{graphicx}
\usepackage[utf8]{inputenc}
 \usepackage[T1]{fontenc}
 \usepackage{amsmath}
 \usepackage{pdfpages}
 \usepackage{amsfonts}
 \usepackage{amssymb}
 \usepackage{makeidx}
 \usepackage{graphicx}
 \usepackage{caption}
 \usepackage{cancel}
 \usepackage{slashed}
 \usepackage{systeme}
 \usepackage{caption}
 \usepackage{wasysym}
  \usepackage{float}
  \usepackage{tikz}
  \usepackage{setspace}
  \usepackage[hidelinks]{hyperref}
  \hypersetup{colorlinks=true,linkcolor=blue,urlcolor=blue}
\begin{document}
\title{Weinberg's 2-factor from helicity constraints }
\author{Fanomezantsoa A. Andriantsarafara$^1$, Andriniaina N. Rasoanaivo$^2$ }

\address{$^1$Physique des Hautes Energies, Faculté des Sciences  BP 906, Université d'Antananarivo, Madagascar}

\address{$^2$Sciences Expérimentales et des Mathématiques, 
Ecole Normale Supérieure d'Antananarivo, \\ Complexe Scolaire Ampefiloha BP 881, Université d'Antananarivo, Madagascar}

\ead{arlivahandriantsarafara@gmail.com}

\begin{abstract}
Scattering amplitudes connect theoretical descriptions to experimental predictions. Low energy terms of the scattering amplitude tend to factorize from the high energy. Different methods have already been established to understand the mechanism of such factorization, Weinberg's theorem. With regard to the Weinberg soft factor, calculations have already shown that this factor has an universal property. In this talk, we show that it is possible to calculate this factor independently from the scattering amplitude based on the Wigner constraints. We also show that such constraints lead us to a system of partial differential equation that simplifies the construction of the Weinberg's soft factor for the case of one particle or two particles. 
\end{abstract}

\section{Introduction}
\paragraph{}In quantum field theory, soft factorization of scattering amplitudes is described by the theorem
of Weinberg \cite{Weinberg:1965nx}. 
In this factorization, the contribution of the soft particles,
that have low energies tending to zero $\left(p_i \to 0 \right) $, factorized out from the high energy part of the scattering amplitude. Such a theorem is also used to show that the effect of attaching a low energy particle to a scattering amplitude is equivalent to multiplying a soft factor to the amplitude, also known as the Weinberg factors. With regard to the in-depth understanding of the factorization, numerous works have been carried out in which such behaviors similarity can be observed in a variety of theories for the low energy  limit of massless particles such as photon \cite{Low:1958sn,Ozeren:2005mp,Hirai:2018ijc}, gluon  \cite{Marotta:2019cip,Campiglia:2017dpg}, and gravitons \cite{Weinberg:1965nx,Cachazo:2014fwa,Pasterski:2015tva}.

Numerous studies have been able to demonstrate that the Weinberg factor is universal \cite{AtulBhatkar:2018kfi,Chakrabarti:2017ltl,Klose:2015xoa}  i.e. it does not depend on the amplitude in question. So the goal of our study is to investigate the universality of the factorization based on Wigner helicity constraints for the case of a single and a double soft limits of gluons.

\section{Weinberg's theorem for a single soft particle}
\subsection{Weinberg's theorem}
\paragraph{}The Figure \ref{figure_soft} is an illustration of the Weinberg's soft theorem in which the low energy behavior is factorized  from the scattering amplitude. The low energy part are known to be a divergent term (IR) and it is called Weinberg factor \cite{Weinberg:1965nx}. Technically, in a single soft limit where one particle goes soft the Weinberg's theorem is written as 
      \begin{equation}\label{theoreme}
      \lim\limits_{s\rightarrow 0}A_{n}\left(1...a\, s\, b...n \right)=S\left(a,s^{\pm},b \right)A_{n-1}\left(1...a\, b...n \right) ,  
      \end{equation}
where $s$ is the label of the the soft particle with low energy and $S$ is the Weinberg's soft factor. In terms of a diagram, Weinberg's theorem is as follows :
\begin{figure}[h!]
\centering
\includegraphics[scale=0.50]{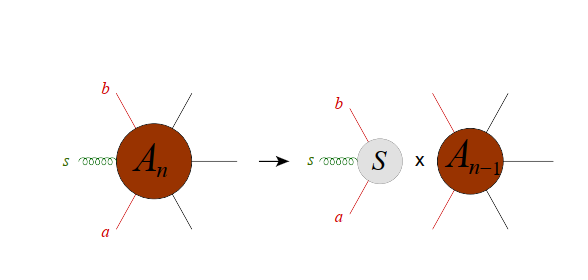}
\captionof{figure}{Weinberg one particle soft theorem.}
\label{figure_soft}
\end{figure}

As shown in \cite{Dixon:2013uaa}, the Weinberg's factors are universal and according to the helicity of the soft particle $s$ they are given by
\begin{equation}\label{sof}
\begin{tabular}{ll}
$S\left(a,s^{+},b \right)=\dfrac{\left\langle a\,b \right\rangle  }{\left\langle a\,s \right\rangle \left\langle s\,b \right\rangle}$, & $S\left(a,s^{-},b\right)=\dfrac{\left[a b \right] }{\left[a s \right]\left[s b \right]} $
\end{tabular}
\end{equation}
where $a$ and $b$ are the adjacent particles to the soft particle.
\paragraph{}The universality of Weinberg's theorem is interpreted by the possibility to calculate the soft factor without independently from the scattering amplitude in question. This factor depends only on the helicity of the low-energy particle and supported by the kinematic variables of its adjacent particles.
\subsection{Helicity constraints}
\paragraph{}The Wigner transformation of the amplitude \cite{Conde:2016vxs} leads to the helicity constraints of the soft factor, see \cite{Rasoanaivo:2020yii}. Such constraints are represented in terms of the  commutation relations between the helicity operator $H_{i}$ of the $i$-th particle and the Weinberg factor $S_{j}$ of the soft $j$-th particle with helicity $h_j$
\begin{equation}\label{sys}
\left[ H_{i},S_{j}\right]=h_{i}S_{j}\delta_{ij}.
\end{equation} 
Our task is to use these constraints in order to fix the kinematics of the Weinberg's soft factor.
An essential ingredient to the resolution of these constraints is the mass dimension of the soft factor. In fact the mass dimension of a $n$-point amplitude $A_n$ of gluons is $\left[A_n \right]= n-4$ which leads to the mass dimension of the single soft factor 
\begin{equation}
\left[S \right]=-1.
\end{equation}

\subsection{With spinor helicity variables}
\paragraph{}In order to simplify the resolution of the helicity constraints it is necessary to use the spinor helicity variables. The spinor helicity variables was introduced to simplify calculations of the scattering amplitudes of massless particles \cite{Dixon:2013uaa}. For the case of radiation of gluons or photons which are massless, in the Weinberg theorem, it is natural to work with these spinor variables also known as the spinor formalism. This formalism is based on the isomorphism between the Lorentz group of transformations $\mathbb{R}^{1+3}$ and the group of invertible square matrices of $SL(2,\mathbb{C}) \otimes SL(2,\mathbb{C})$ on spinors. In other words, for each Lorentz vector element of $\mathbb{R}^{1+3}$ corresponds to a bispinor element of $SL(2, \mathbb{C}) \otimes SL( 2, \mathbb{C})$ through
\begin{equation}\label{d}
p_\mu\longrightarrow p_{a\dot{b}}=\lambda_{a}\tilde{\lambda}_{\dot{b}}
,				
\end{equation}
where $\lambda$ and $\bar\lambda$  are respectively a left and a right handed spinors. With these new variables we can define two new invariant products first a product between the left handed spinors denoted by $\langle ij\rangle$ and the second a product between the right handed spinors denoted by $[ij]$. In terms of the spinor variable the system \eqref{sys} can be written as 
\begin{equation}\label{pde}
-\frac{1}{2}\left (\lambda_j^a\frac{\partial}{\partial \lambda_j^a}-\bar{\lambda}_j^{\dot{a}}\frac{\partial}{\partial \bar{\lambda}_j^{\dot{a}}}\right ) S_j= \begin{cases}
h_{j}S_j &\text{if } \quad i=j \\
0 &\text{otherwise}
\end{cases}
\end{equation}

For the case where the particle 1 is the soft particle. The Weinberg factor only depends on its helicity and on the kinematics of its neighbors $n$ and $2$. The resolution of this system of partial differential equations through separation of variables with the  mass dimension as a boundary condition, we obtain the Weinberg factors depending on the helicity $h_j$ of the soft particle 
 \begin{equation}
             \begin{tabular}{lcr}
             $S\left( 1^{+}\right) =\dfrac{\left\langle n 2 \right\rangle }{\left\langle n 1 \right\rangle\left\langle12 \right\rangle},$ & &$S\left( 1^{-}\right) =\dfrac{\left[ n2\right] }{\left[ n1\right]\left[12 \right]  }.$
             \end{tabular}
            \end{equation} 
\section{Weinberg's two-particle theorem}

\paragraph{}For the case of double soft theorem it is necessary to review the way how we perform the soft limits. We can either taking a successive limits or taking a simultaneous limit of the two soft particles.

In the successive soft limits, we first perform one soft limit on the scattering amplitude then applying a second soft limit. For a universal property of the soft theorem such operation should lead to the same result even we invert the order of the limits. Let us consider $1$ and $2$ to be the soft particles, in which the consecutive limits are 
\begin{align}
& A_{n} \xrightarrow{\quad 1\to 0\quad} \;S_{1}A_{n-1}\xrightarrow{\quad 2\to 0\quad} \; S_{1}S_{2}A_{n-2}\\
& A_{n} \xrightarrow{\quad 2\to 0\quad} \;S_{2}A_{n-1}\xrightarrow{\quad 1\to 0\quad} \; S_{2}S_{1}A_{n-2}
\end{align} 
In general $S_{1}S_{2}$ and $ S_{2}S_{1}$ are not the equal, particularly in the case where the helicity $h$ of the two soft particles are different. Solving this problem demands a better way to think of the double soft limit which is a simultaneous soft limit. For that we will use the system \eqref{sys} for the case of double soft and with the sipinor variables we convert it into a system of PDE version of the Wigner constraints.

\paragraph{}In the same way as in \eqref{pde}, the simultaneous double limit is reduced to a set of Wigner constraints which is a system of partial differential equation where the low energy particles are $i=1,2 $ with the respective helicity $h_{1}=+1$ and $h_{2}=-1$. The solution of the corresponding PDE leads to
\begin{equation}\label{xx}
S=\dfrac{\left\langle 1 3 \right\rangle^{\alpha_{1}+h_{1}+\beta_{1}+\gamma_{1}} \left\langle 2 3 \right\rangle^{\alpha_{2}+h_{2}+\beta_{1}+\gamma_{2}} \left\langle 3n \right\rangle^{\sigma} }{\left\langle 1 n \right\rangle^{\gamma_{1}}\left\langle 2 n\right\rangle^{\gamma_{2}} \left\langle 1 2 \right\rangle^{\beta_{1}} }\dfrac{\left[ 1 3 \right]^{\alpha_{1}-h_{1}+\tilde{\beta_{1}}+\tilde{\gamma_{1}}} \left[ 2 3 \right]^{\alpha_{2}-h_{2}+\tilde{\beta_{1}}+\tilde{\gamma_{2}}} \left[ 3 n\right]^{\sigma}  }{\left[1 n \right]^{\tilde{\gamma_{1}}}\left[ 2 n\right]^{\tilde{\gamma_{2}}} \left[ 1 2 \right]^{\tilde{\beta_{1}}}  } \; .
\end{equation}
where $\alpha_{1},\alpha_{2}, \beta_{1},\;\beta_{2},\;\gamma_{1},\;\gamma_{2},\;\sigma,\;\tilde{\beta_{1}},\; \tilde{\gamma_{1}},\; \tilde{\gamma_{2}} $ are parameters from the separation of variables to reduce the PDE into independent ordinary differential equations. Moreover, to determine these parameters we use the mass dimension $[S(1,2)]=-2$ and we also recall  the collinear behavior of the double soft as $z=\left\langle 12 \right\rangle \to 0$ and $\tilde{z}=\left[ 12\right] \to 0$. The IR from the collinear limits leads to the fact that $S $ must be a multiple of $\dfrac{1}{z}$ or $\dfrac{1}{\tilde{z}}$. This implies the Weinberg 2-factor must is necessarily the linear combination of the Weinberg 1-factor of the two successive limits of the soft particles. We can then write that 
\begin{equation}
S\left( 1,2\right)=f\,S_{1}S_{2}+g\,S_{2}S_{1}, 
\end{equation} 
where $f$ and $g$ are functions that doesn't transform under a Wigner transformation.
\paragraph{} To determine $f$ and $g$, we need to taken into account that for ordered double soft limit $E_{2}\ll E_{1}$, that is to say that particle 1 is softer than particle 2, the Weinberg factor of \eqref{xx} should tend to the successive soft factor $S\left( 1\right) S\left( 2\right) $ and in case $E_{1}\ll E_{2}$, the Weinberg factor of \eqref{xx} should tend to $S\left( 2\right)S\left( 1\right)$.
In other words :
\begin{itemize}
\item[$\bullet$]  If the particle 2 is softer than the particle 1, the value that $g$ should tend to 1 and  $f$ tends to zero.
\item [$\bullet$] In the other case where the particle 1 is softer than the particle 2, $g$ should tend to zero and $f$ tends to 1.
\end{itemize}
Such functions should be given by the following
    \begin{equation}
      \begin{tabular}{lcl}
      $f =\left( \dfrac{k_{1}\cdot q}{\left(k_{1}+k_{2} \right)\cdot q }\right)^{\alpha} $ & and &$g =\left( \dfrac{k_{2}\cdot q}{\left(k_{1}+k_{2} \right)\cdot q }\right)^{\alpha} ,$
      \end{tabular}
     \end{equation}
where $q$ is an arbitrary momentum; and based on the solution in \cite{Klose:2015xoa} the value of $\alpha$ is equal to 1. We can then write that the two-particle Weinberg factor takes the following form :
 \begin{equation}\label{vita}
 S(1,2)=\dfrac{1}{\left\langle n\left( 1+2\right)3 \right] }\left( \dfrac{1}{\left\langle n2\right\rangle\left[ n2\right] } \dfrac{\left[ n3\right]\left\langle n2\right\rangle^{3} }{ \left\langle n1\right\rangle \left\langle 12\right\rangle }- \dfrac{1}{\left\langle 32\right\rangle \left[ 32\right] }\dfrac{\left\langle n2\right\rangle \left[ 31\right] ^{3}}{\left[ 12\right] \left[ 23\right] }\right).
 \end{equation}           
\section{Conclusion and Discussion}
 \paragraph{}In this work, we investigate the Weinberg factor using the Wigner constraints which is the same as solving a system of partial differential equations with the spinor variables. For the case of particles where the helicity take the Wigner's constraints are linear and only depends on the momenta of the adjacent particles. For the case of double soft limit with the consecutive limit, we discuss that the order of the limit as an effect on the Weinberg's 2-factor of the soft particles. The resolution of the Wigner's constraints allows us to fix this problem and leads to a linear combination of the consecutive limits. In the resolution we didn't find enough argument for the choice of $\alpha=1$ in the very last step of the resolution. In our case it was still a free parameter, so in the future we expect to find the physics behind this parameter that makes it equal to 1. 
 \section*{Acknowledgement}
\paragraph{} This work would have been impossible without the support of the High Energy Physics (PHE) of the sciences faculty of the university of Antananarivo. I also thank Andriniaina Narindra Rasoanaivo since without him this work would not have taken place. I would also like to thank my colleagues that contributed directly or indirectly to the development of this work.
 
\end{document}